\documentstyle[aps,12pt,preprint,tighten]{revtex}

\begin{document}
\draft
\title{High-${T_c}$ Superconductivity of  van der Waals Condensates}
\author{V. N. Bogomolov}
\address
{\small A.F.Ioffe Physicotechnical Institute, Russian Academy of Sciences,\\
St. Petersburg, Russia\\
e-mail V.Bogomolov@shuvpop.ioffe.rssi.ru}
\normalsize
\date{}
\maketitle
\begin{abstract}
 The paper considers the possibility of superconductivity setting
in when the distance $a_B$ between molecules (atoms) is such that
$a_F < a_B < a_W$, where $a_F$ and $a_W$ are the equilibrium separations in
Fermi metals and van der Waals insulators. The last has no band sructure.
This is the binding energy region 40--60 kJ/mol. However, neither stretch
of metals ("dilution" by insulators) nor compression of insulators to
$a = a_B$ produce substances stable under normal conditions. It was proposed
to stabilize such substances prepared of insulators by adsorbing their
atoms (molecules) on metal catalysts to obtain (Metal)(Insulator)$_X$-type
systems for $X > 1$. The well-known method of metal sputtering in a gaseous
atmosphere has used to successfully obtain such "sorption compounds" and
collect with magnetic funnels clusters of diamagnetic particles kept
suspended in a magnetic field gradient $\sim$100 G/cm  at 300 K
(the levitation effect).
\end{abstract}

\section{Introduction}
In the 1930s, at least two significant steps in the physics of condensed
state were made with due account of the reality of localized states.
This is the concept of the polarons and the explanation of the
van der Waals forces in molecular crystals (MC), which involved considering
the electron in this case as a particle. Taking into account the
corpuscular properties of particles and of quasi-particles is of
fundamental importance for an adequate description of a system [1].

New information on the properties of MC subjected to extreme conditions
suggests apparently that a more adequate description of MC properties
requires the inclusion of other well localized electronic states as well.

\section{Description of the MC structure}  The physical properties of MC
made up of molecules (atoms) with filled electronic shells are usually
described by band theory. The equilibrium
distances $a_W$ in MC are close to twice the radii of the excited state
orbitals $r_1$. They are, however, several times larger than the double
ground-state orbital radii. For instance, in the xenon condensate
$aW/2 \sim r_1 = 2.2$ \AA, whereas $r_0 = 0.6 $ \AA. In condensates of
atoms with unfilled electronic shells (fi, metals) $a_F/2 \sim r_0$.
Both the binding energies and the electrical conductivity of the MC
are extremely small.

  Polaron crystals can exist, besides the band, in the semiclassical
regime too (the small polarons). In this case, the locality is originates
from electron interactions with the "heavy" medium. A similar phenomenon
can be conceived to exist in MC too, if one takes into account the locality
of the pair excitation of neughboring molecules M (atoms A) and
the attendant formation of the virtual molecules M$_2^*$ (or A$_2^*$).
At low concentrations, the electron pairs binding each such heavy
double-center molecule may behave semiclassically, as the polaron electron
does.

 In 1938, F. London suggested that superconductivity might be associated
with the
Bose--Einstein condensation in systems where fermions could be replaced
by bosons [2]. In 1946, Ogg, drawing on [2], prepared from metal-ammonia
solutions a substance "...metastable in the "forbidden" concentration region,
which is, thus, characterized by Bose--Einstein condensation of trapped
electron pairs" [3].

 A feeling of the band theory being inadequate to a description of the
MC properties in unusual, extreme physical conditions has appeared
in 1989, when high-precision measurements of the optical properties of
xenon were performed at pressures up to 2 Mbar, and their interpretation
in terms of band theory was attempted [4,5]. By that time, experimental
studies of the interaction of the metal--insulator- and
insulator--insulator-type in atomic-scale zeolite channels had been
carried out, and scenarios of local mechanisms of wetting, adsorption,
and catalysis were developed [6]. This permitted one to describe the
pressure-induced MC metallization as resulting not from band overlap
but rather from overcoming the percolation threshold (the formation
of an infinite new-phase cluster) in a "gas" of localized and independent
charges at lattice sites with increasing charge concentration [7].

The model of the MC based on the concept of well localized pair excitations
is actually simple [6-10]. The Xenon crystal, for instance (a lattice of Xe
atoms at van der Waals distances $a_W\sim 4.4 $ \AA), is considered as
a mixture (in the classical sense) of atoms in the ground and excited states.
At each instant of time, $\sim$80\% of atoms are in the ground state
($r_0 =$ 0.6 \AA) and are slowly attracted to one another by the
van der Waals forces. The remaining 20\% of atoms are maintained in
excited state by random pair interactions ($r_1=2.2$ \AA) and form
noninteracting, covalently bound virtual molecules Xe$_2^*$ scattered
over the lattice (an analog of Cs$_2$ molecules with participation
of two $6s$ electrons). It is these molecules that determine the equilibrium
lattice parameter $a_W\sim 4.4$ \AA. Virtual Xe$_2^*$ molecules were
experimentally detected in optical measurements in 1975 in studies
of  Cs and Xe mixtures [11]. In normal conditions, the concentration of
the Xe$_2^*$ molecules is small, their separation $R>>a_W$, they
form a disordered system and do not interact with one another in the way
that could give rise to a real electronic band. Compression increases
the concentration of the Xe$_2^*$ molecules having two paired $s$-electrons.
Such double-center bosons exist, similar to Cooper pairs, due to interaction
with the heavy atomic-core lattice. Compression increases not only the
concentration of the Xe$_2^*$ molecules, but the degree of their ordering
as well, because they occupy regular lattice sites. The conditions in this
nonequilibrium concentration region of electronic molecular pairs may
become favorable for their Bose condensation before the system transforms
to a Fermi metal. Thus,  van der Waals MC ($a = a_W$, molecules or atoms
with filled electronic shells)  and Fermi systems ($a = a_F$, atoms with
nonfilled electronic shells) may be separated by a region of Bose
superconductors ($a = a_B$). In other words,
$a_W\sim 2r_1 > a_B > a_F\sim 2r_0$ [9]. The Xe$_2^*$ binding energies
$Q^* = kT_c >> q_W$, which is the binding energy of the MC lattice.
In the case of Xe, these are $\sim$0.5 and 0.13 eV, respectively.

For a large number of MC, $q_W < $ 40 kJ/mol, and for metals,
$q_F > 60$ kJ/mol [9,10]. The binding energy of metallized (and
superconducting? [8]) Xenon, $q_B\sim 50$ kJ/mol $\sim Q^*=kT_c\sim$ 0.5 eV.
If we relate this energy with the superconducting state, this
superconductivity is superhigh-temperature ($T_c\sim$ 5000 K). The
conclusion of the existence between the
insulators and metals of a region of superconductors was reached
also in [12]. The attempt made in 2000 to reveal the Xenon superconductivity
has not met with success, because one could not reach the region of reliable
metallization [13].

\section{Experiments} There are more than one method to keep a system in
the "forbidden" concentration region. Among them are, for instance,
methods involving chemical interactions, freezing of solutions [3],
or making use of adsorption forces in (Metal)(Nonmetal)$_X$-type
"sorption compounds" [9,10]. We used for the preparation of such
nanocomposites the well-known techniques involving metal evaporation
in a gaseous atmosphere. There are presently no known stable
(Metal)(Nonmetal)$_X$-type compounds ($X>1$). A simple way to
investigate the physical properties of such compounds lies in probing
(Metal)(Nonmetal)$_X$ clusters before they disintegrate on the substrate.
This method was employed to measure the CaXe cluster optical-bond
energy ($\sim$0.2 eV) [14]. The bond energies of PdXe and PtXe were
calculated to be 0.42 and 1.0 eV, respectively [15]. The adsorption
energy of Xe atoms on Pd surface was found experimentally to be close
to 0.3--0.4 eV [16].

To separate diamagnetic clusters and their aggregates prepared by this
method, they were passed through a nonuniform magnetic field before
they disintegrated in collision with the substrate, and collected
in magnetic funnels. A compact black cloud of diamagnetic
particles was suspended in a magnetic funnel (levitation in a magnetic
field gradient of about 100 G/cm at 300 K). A detailed description of
these experiments is being presently prepared for publication.

\vskip0.5cm
The properties of metallized  van der Waals condensates may thus
far be a manifestation of superconductivity with $T_C>$ 300 K.

\section*{References}
\begin{enumerate}
\item   E.Wigner. Symmetries and Reflections.  Blumington-London (1970), p.
78.
\item   F.London. Phys. Rev. {\bf54}, 947 (1938).
\item   R.A.Ogg . jr. Phys. Rev.  {\bf69},  243 (1946).
\item   ë.á.Goettel, J.H.Eggert, I.F.Silvera, W.C.Moss.
Phys. Rev. Lett. {\bf62}, 665 (1989).
\item   R.R.Reichlin, K.F.Brister, A.K.McMahan, M.Ross, S.Martin, Y.K.Vohra,
A.L.Ruof. Phys. Rev. Lett.  {\bf62}, 669 (1989).
\item   V.N.Bogomolov. Phys. Rev. {\bf51}, 17040 (1995).
\item   V.N.Bogomolov. Tech. Phys. Lett.  {\bf21}, 928 (1995).
\item   V.N.Bogomolov. Metallic Xenon.  Conductivity or Superconductivity?
Preprint 1734,  Russian Academy of Sciences,  A.F.Ioffe Physical
Technical Institute (1999).
\item   V.N.Bogomolov.  cond-mat/9912043.
\item  V.N.Bogomolov.  cond-mat/0103099.
\item  R.A.Tilton, C.P.Flynn.  Phys. Rev. Lett. {\bf34}, 20 (1975).
\item  M.Capone, M.Fabrizio, E.Tossatti.  cond-mat/0101402
\item  M.I.Eremets, E.A.Gregoryanz, V.V.Struzhkin,  H.Mao,  R.J.Hemley,
N.Mulders, N.M.Zimmerman. Phys. Rev. Lett. {\bf85},  2797  (2000).
\item  A.W.K.Lleung, J.G.Kaup, D.Bellert, J.G.McGaffrey, W.H.Breckenridge.
J. Chem. Phys. {\bf111},  981 (1999).
\item  J.V.Burda, N.Runeberg, P.Puukko. Chem. Phys. Lett. {\bf288}, 635
(1998).
\item  A.Zangwill.  Physics of Surfaces. (Camb. Univ. Press,
Cambridge) (1988) p. 256.
\end{enumerate}
\end{document}